\documentclass[prb,aps,showpacs,twocolumn,longbibliography]{revtex4-1}
\usepackage{times}
\usepackage{graphicx}
\usepackage{float}
\usepackage{latexsym,amsmath,amssymb,bm,euscript}
\usepackage{color}
\usepackage{epstopdf}
\usepackage[colorlinks=true,linkcolor=blue,citecolor=blue,urlcolor=blue]{hyperref}
\usepackage{hyperref}

\begin{document}

\title{Fragility of the nematic spin liquid induced by diagonal couplings in the square-lattice SU(3) model}
\author{Xiao-Tian Zhang$^{1}$}
\author{Wen-Jun Hu$^{2}$}
\author{Elbio Dagotto$^{2,3}$}
\author{Shoushu Gong$^1$}
\affiliation{
$^1$ Department of Physics, Beihang University, Beijing 100191, China\\
$^2$ Department of Physics and Astronomy, University of Tennessee, Knoxville, Tennessee 37996, USA\\
$^3$ Materials Science and Technology Division, Oak Ridge National Laboratory, Oak Ridge, Tennessee 37831, USA
}

\begin{abstract}
We present a large-scale density matrix renormalization group (DMRG) study of the spin-$1$ SU(3) bilinear-biquadratic model on the square lattice, which was suggested to host a nematic spin liquid state in recent DMRG calculations. We report that this spin liquid appears to strongly compete with a three-sublattice magnetic order. To further study the competition between the two states, and the reason of the emergent nematic spin liquid, we included an additional next-nearest-neighbor SU(3) symmetric interactions along {\it one} of the two plaquette diagonal directions. This allows to tune the square-lattice model to the triangular-lattice model. By computing spin correlation functions and various order parameters, we find that the three-sublattice order may develop at {\it infinitesimal} additional new coupling, at least within the precision of our study. Compared with the previous findings that the nematic spin liquid is stable in extended parameter regions with additional couplings that respect the lattice symmetries of the square lattice, we argue that here the diagonal couplings, which frustrate the bipartite-lattice structure, rapidly suppress the two-sublattice fluctuations and the three-sublattice order thus wins. This numerical result is consistent with the conjecture that the nematic spin liquid emerges from the competition between two- and three-sublattice fluctuations.
\end{abstract}

\maketitle

\section{Introduction}

Frustrated magnets are one of the most active and challenging directions in modern condensed matter physics. 
Various exotic quantum phases have been found in spin systems with frustration incorporated, among which the quantum spin liquids (QSLs) are the most interesting ones. 
QSLs host disordered phases even down to zero temperature, and even more remarkably, they have long-range entanglement~\cite{wen1990, chen2010} as well as fractionalized excitations~\cite{Wen1991, Senthil2000, Senthil2001, HHLai2020}. 
Because of these exotic properties, QSLs have been searched for extensively, and recently spin-liquid-like phases have been reported in several spin $S=1/2$ antiferromagnetic materials and spin liquid states have been found in some frustrated models by unbiased numerical simulations~\cite{savary2016, zhou2017}.

The search for novel quantum phases has also been extended to spin $S=1$ systems both experimentally~\cite{nakatsuji2005,cheng2011,fak2017,quilliam2016,cava2019} and theoretically~\cite{Haldane1983_2, aklt1987, katsumata1989, hagiwara1990, white1993, schollwock1996, shelton1996, lauchli2006, corboz2007, corboz2017, chen2018, chen2020, boos2020, WJHu2020, YXHuang2021}. 
There are several spin-$1$ materials that have been reported to host spin-liquid ground states, such as the triangular layered materials NiGa$_2$S$_4$~\cite{nakatsuji2005}, Ba$_3$NiSb$_2$O$_9$~\cite{cheng2011, fak2017}, and the layered honeycomb-lattice material 6HB-Ba$_3$NiSb$_2$O$_9$~\cite{quilliam2016}. 

Theoretically, the bilinear-biquadratic Hamiltonian has been suggested as a prototypical model for describing a spin-$1$ system beyond the Heisenberg model~\cite{papanicolaou1988}. This extended model is defined as
\begin{equation}\label{model}
H = \sum_{i,j} J_{ij} {\bf S}_i \cdot {\bf S}_j + K_{ij} ( {\bf S}_i \cdot {\bf S}_j )^2,
\end{equation}
where ${\bf S}_{i}$ is the spin-$1$ operator at site $i$, and $J_{ij}$ and $K_{ij}$ are the bilinear and biquadratic interactions, respectively. 
In this model, the highly symmetric SU(3) point with equal bilinear and biquadratic interactions $J_{ij}=K_{ij}$ is particularly interesting. 
The strong competition between the two couplings, together with the SU(3) symmetry, enhances the frustration, which may lead to novel quantum phases. 
An exact diagonalization study found a rich phase diagram for the one-dimensional spin-$1$ bilinear-biquadratic model with nearest-neighbor (NN) couplings~\cite{lauchli2006}, unveiling a quantum phase transition near the SU(3) point, between a three-sublattice quadrupolar order and the gapped Haldane phase~\cite{Haldane1983_1, Haldane1983_2}.

In two dimensions, intensive theoretical studies have focused on the highly symmetric SU(3) point of the model Eq.~\eqref{model}~\cite{changlani,lauchli2006_2,toth2010, toth2012,bauer2012,zhao2012,corboz2012,corboz2013,corboz2017,niesen2017tensor,corboz2018,hunsl}. 
By considering only the NN couplings, a trimerized valence bond state~\cite{changlani,corboz2012} and a plaquette valence bond state~\cite{changlani,corboz2012} have been found as the ground states on the kagome and honeycomb lattices, respectively. 
These emergent states that spontaneously break the translational symmetry of the lattice, rather than the magnetic ordering, indicate that the three-flavor degree of freedom is frustrated by these lattice geometries.
On the other hand, on the triangular lattice the use of linear flavor wave theory found a three-sublattice magnetic order with a finite order moment~\cite{lauchli2006_2,bauer2012}. This result was also confirmed by the density matrix renormalization group (DMRG)~\cite{bauer2012} and the infinite projected entangled pair states (iPEPS)~\cite{corboz2018} calculations.

\begin{figure}[t]
\includegraphics[width = 0.8\linewidth]{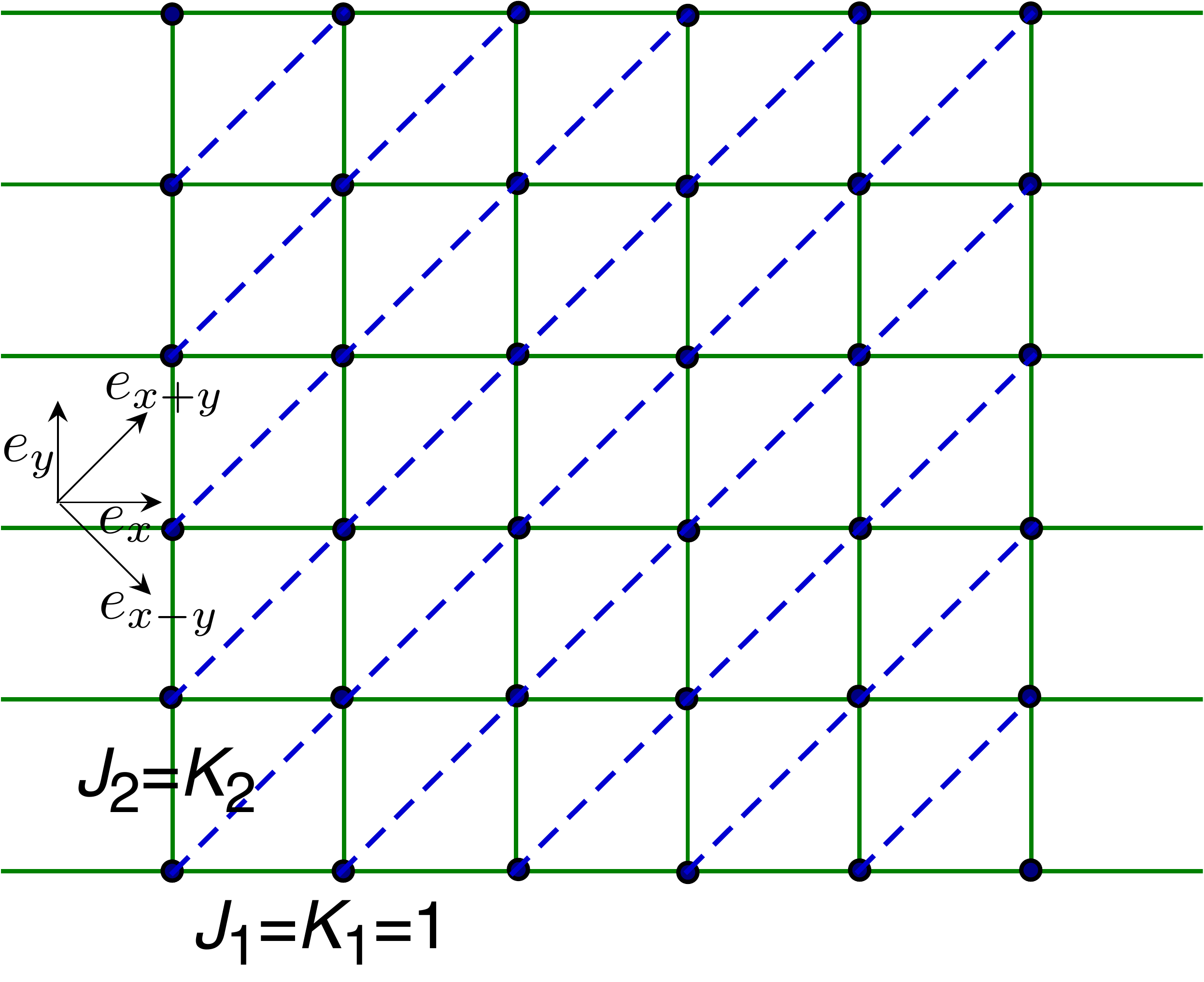}
\caption{Schematic representation of the spin-$1$ SU(3) model studied here.
The model has the previously used nearest-neighbor (NN) $J_1, K_1$ interactions and the new additional
next-nearest-neighbor (NNN) $J_2, K_2$ interactions along one of the diagonal directions $e_{x+y}$. We consider the equal bilinear-biquadratic interactions $J_1 = K_1 = 1$ and $J_2 = K_2$. For $J_2 = K_2 = 1$, the model realizes the isotropic triangular-lattice model. For $J_2 = K_2 = 0$, it reduces to the square-lattice model. Thus, the model interpolates between a triangular and a square lattice. The sketch also illustrates the cylindrical geometry used in the DMRG calculation, which has the periodic boundary conditions along the $y$ direction $e_y$ and the open boundary conditions along the $x$ direction $e_x$.}
\label{lattice}
\end{figure}

The SU(3) model on the square lattice may be the most intriguing system, due to the additional frustration introduced by the competition between two- and three-sublattice fluctuations~\cite{toth2010, toth2012,bauer2012,corboz2017,niesen2017tensor,hunsl}. 
While analytical methods, including linear flavor wave~\cite{toth2010} and Schwinger boson mean-field theories~\cite{bauer2012}, predicted a three-sublattice order, these calculations all failed to obtain a finite order parameter for neither spin dipolar nor quadrupolar degrees of freedom, because the reduction of the order moments always diverges near the SU(3) model due to the presence of gapless lines in the boson dispersions~\cite{bauer2012,hunsl}.
Therefore, advanced quantum computational methods are required to study this system. 
Indeed, previous DMRG calculations on the cylindrical geometry, with shifted boundary conditions (designed to match the three-sublattice structure)~\cite{bauer2012}, and iPEPS simulations~\cite{bauer2012,corboz2018} also suggested a three-sublattice order ground state that is similar to the state on the triangular lattice. 
Interestingly, recent large-scale DMRG calculations on the rectangular cylinders with regular boundary conditions (see Fig.~\ref{lattice}) found consistent evidence for a nematic spin liquid phase emergent around the SU(3) point, which spontaneously breaks the lattice rotational symmetry possibly due to the dominant fluctuations at the momentum ${\bf q} = (\pi, 2\pi/3)$~\cite{hunsl}. 
This dominant momentum may be considered as a result of the strong competition between two- and three-sublattice fluctuations.
Indeed, in the presence of some additional further-neighbor interactions, which respect the lattice symmetries of the square lattice and do not frustrate either the two- or three-sublattice structures, the nematic spin liquid was found to be stable in an extended parameter region~\cite{hunematic}.
However, since the three-sublattice magnetic order has not been clearly identified close to the nematic spin liquid in previous studies~\cite{hunsl,hunematic}, it remains elusive how the two states compete near the SU(3) model and it is unclear whether the nematic spin liquid indeed originates from the competition between two- and three-sublattice fluctuations.

To shed  light on these open questions, we study the square-lattice SU(3) model with additional next-nearest-neighbor (NNN) interactions $J_2 = K_2$ [also SU(3) symmetric] along the $e_{x+y}$ direction (one of the two diagonal directions on the square lattice), as shown in Fig.~\ref{lattice}. 
With growing $J_2, K_2$ couplings, the system evolves from the square-lattice (with the nematic spin-liquid ground state) to the triangular-lattice model (where the ground state has been identified as the three-sublattice order)~\cite{lauchli2006_2,bauer2012,corboz2018}.
In this paper, we perform large-scale DMRG calculations for this model on a rectangular cylindrical geometry. 
Through the calculations of spin correlation functions and different order parameters, we find that the three-sublattice order appears to develop at infinitesimal $J_2 = K_2$ couplings, within the precision of our effort, and the nematic spin liquid becomes, thus, unstable in the presence of these additional diagonal couplings.
 
Compared with the previously considered couplings along both the $e_{x+y}$ and $e_{x-y}$ directions~\cite{hunematic}, here the couplings break the lattice $C_4$ rotational symmetry, as well as the mirror symmetries with respect to either $e_{x}$ or $e_{y}$ direction.
Thus, the bipartite two-sublattice structure would be frustrated.
Our DMRG results suggest that these additional couplings quickly suppress the two-sublattice fluctuations and lead to the emergence of three-sublattice order. This result indicates the crucial role of balancing two- and three-sublattice fluctuations to stabilize the nematic spin liquid state.

The organization of our paper is as follows. 
In Sec.~\ref{sec:method}, we introduce the model and the details of the DMRG simulations. 
In Sec.~\ref{sec:results}, we show our numerical results including the spin correlations and different order parameters. The final section is the summary and discussion in Sec.~\ref{sec:sum}.

\begin{figure}[t]
\includegraphics[width = 1\linewidth]{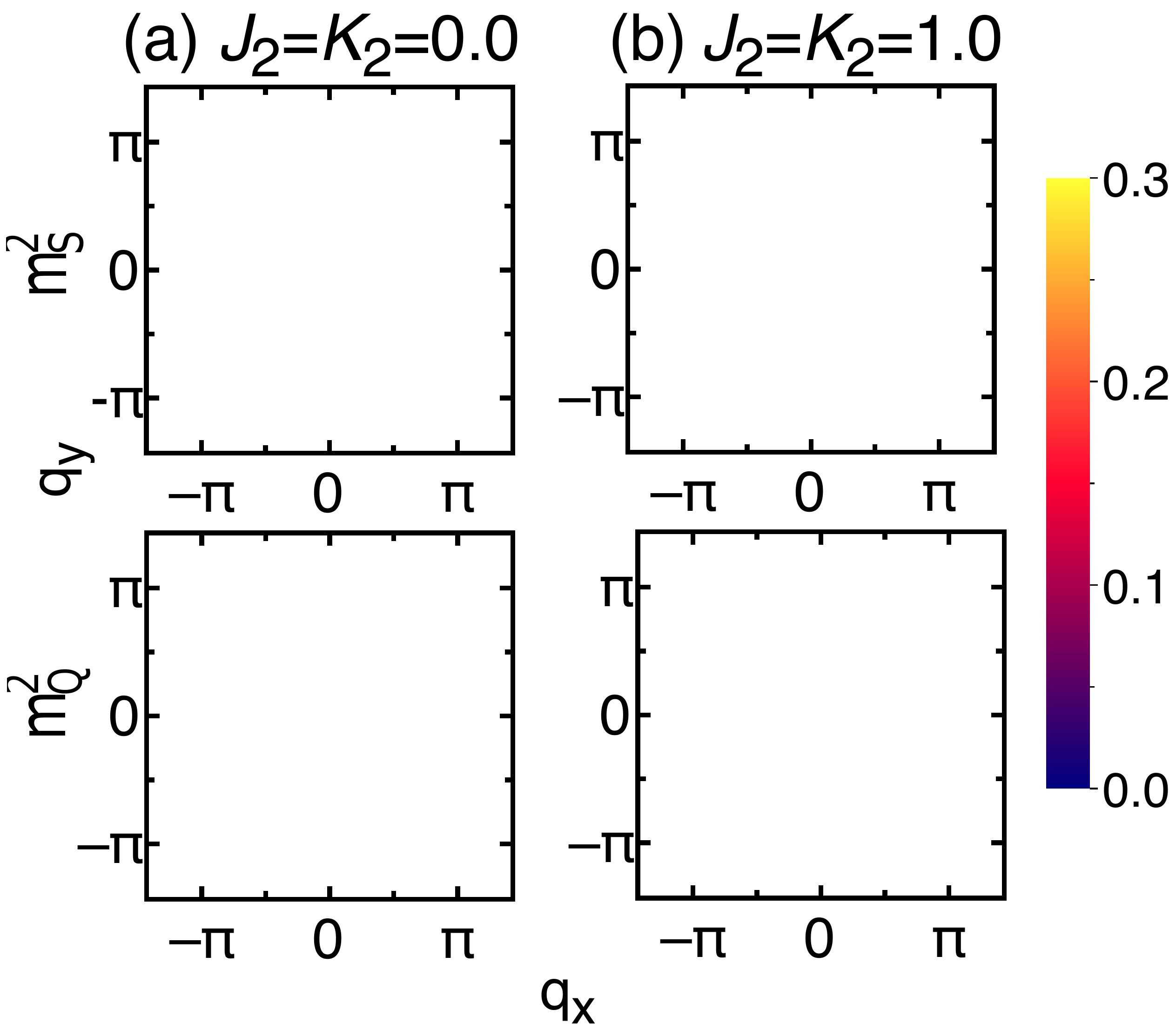}
\caption{The spin dipolar ($m^2_S$) and quadrupolar ($m^2_Q$) order parameters for (a) $J_2 = K_2 = 0$ and (b) $J_2 = K_2 = 1$. Both order parameters are obtained from the corresponding correlation functions of the middle $6 \times 12$ sites on a long RC6 cylinder. While the order parameters are dominated by elongated horizontal regions at ${\bf q} = (\pi, 2\pi/3)$ in the spin liquid state (a), prominent sharp point-like peaks appear at ${\bf q} = (2\pi/3, 2\pi/3)$ in the three-sublattice order (b).}
\label{pd}
\end{figure}

\begin{figure}[!t]
\includegraphics[width = 0.9\linewidth]{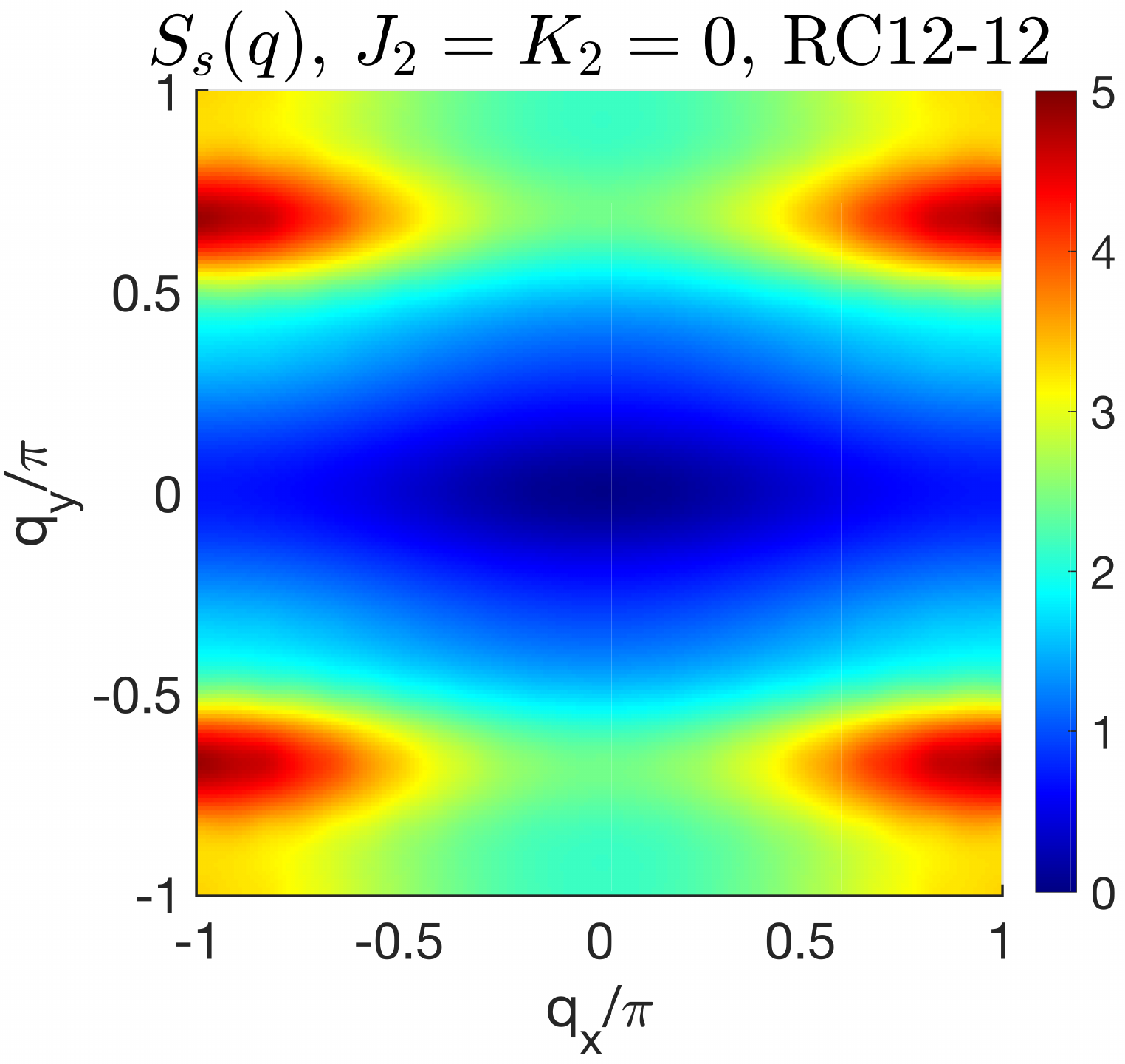}
\caption{Spin structure factor $S_S(\bf q)$ of the SU(3) model with $J_2 = K_2 = 0$ on the RC12-32 cylinder.
The structure factor is computed from the Fourier transform of the spin correlations for the middle $12 \times 12$ sites, which is featureless at ${\bf q} = (2\pi/3, 2\pi/3)$ but shows rounded peaks at ${\bf q} = (\pi, 2\pi/3)$.}
\label{Ms2}
\end{figure}

\begin{figure}[!t]
\includegraphics[width = 0.9\linewidth]{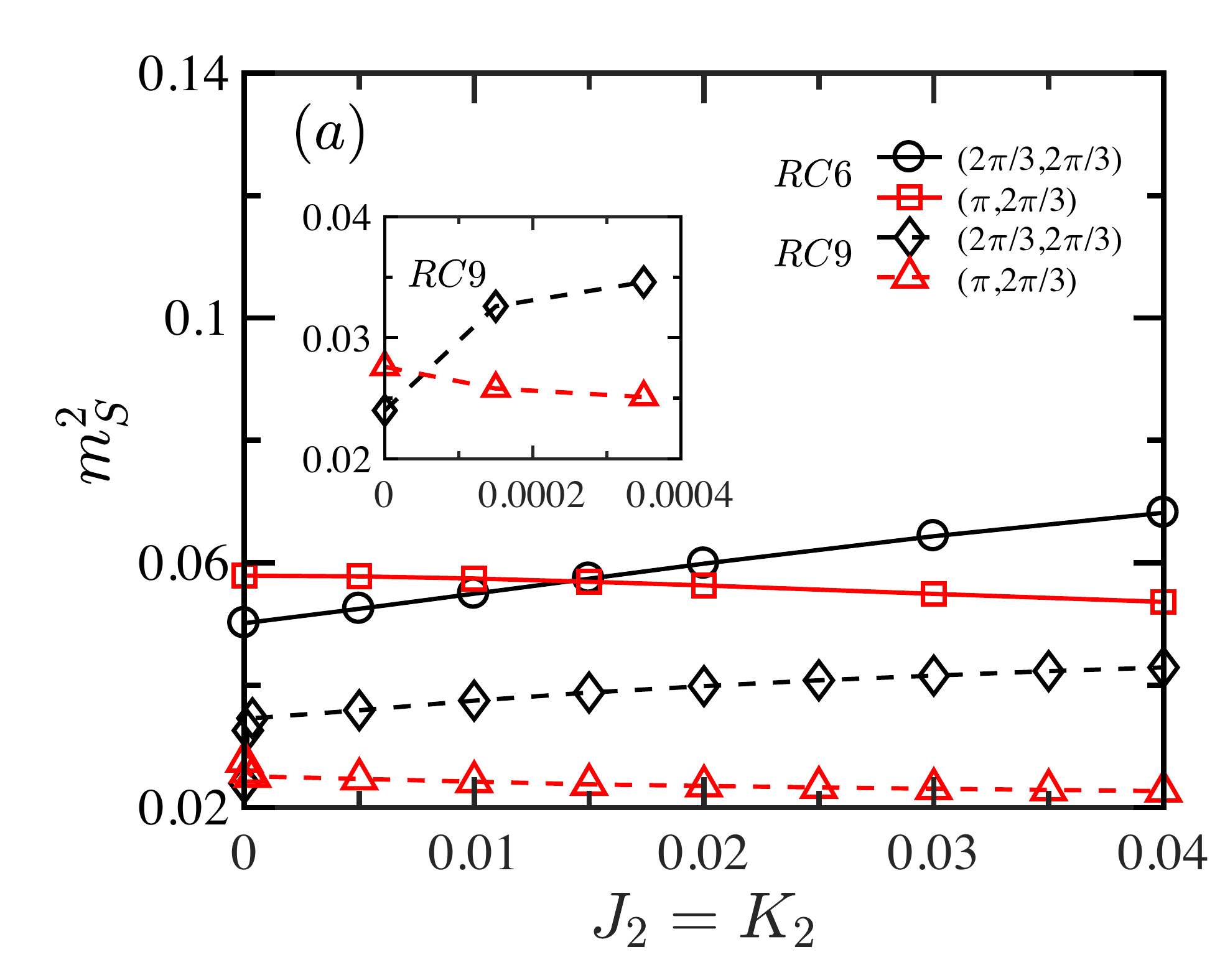}
\includegraphics[width = 0.9\linewidth]{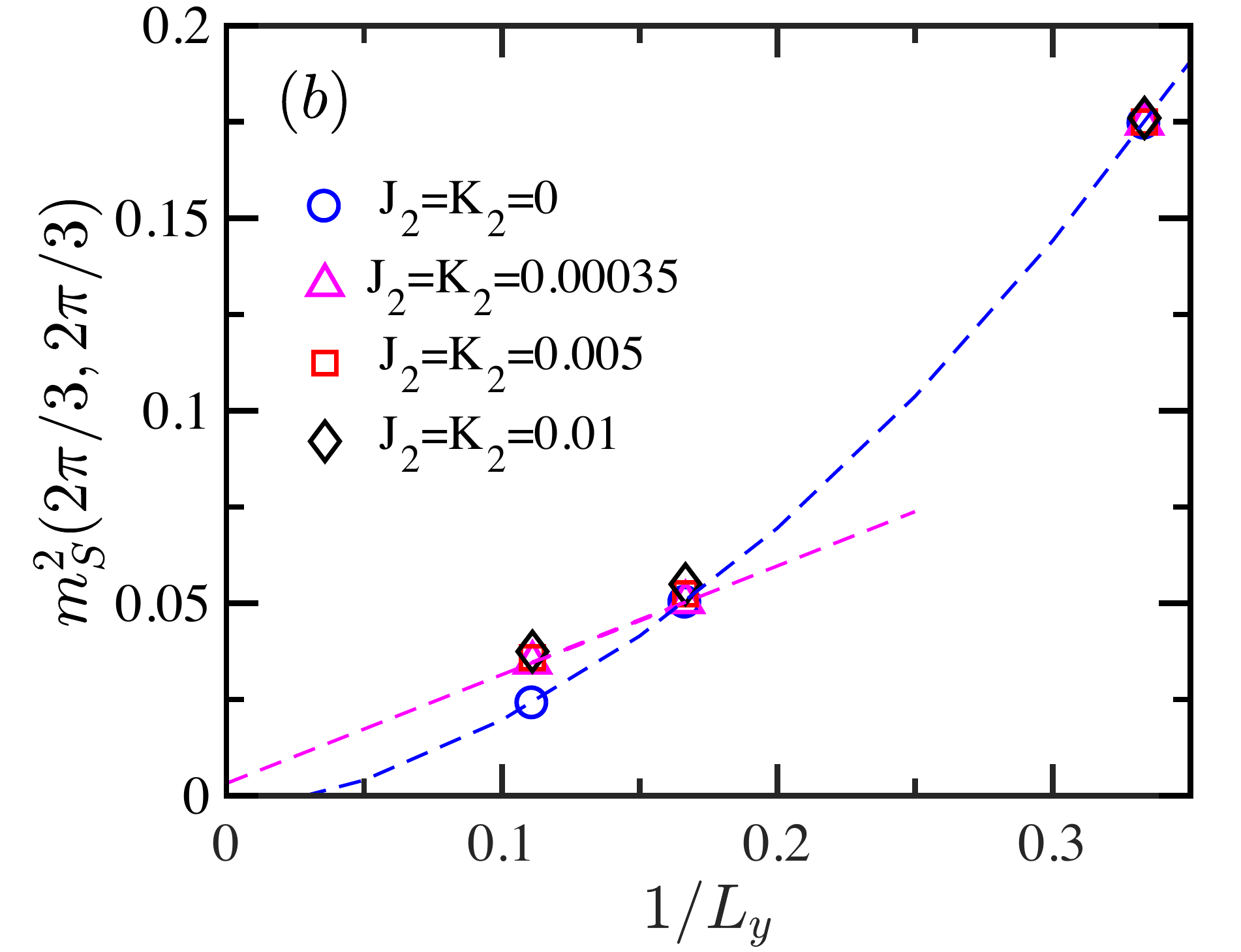}
\caption{Coupling dependence and size scaling of the magnetic dipolar order parameter $m^2_S$. (a) Coupling dependence of $m^2_S$ at ${\bf q} = (2\pi/3, 2\pi/3)$ and $(\pi, 2\pi/3)$ on the RC6 and RC9 cylinders. The inset shows the zoom in of the RC9 data at very small couplings. (b) Finite-size scaling of $m^2_S(2\pi/3, 2\pi/3)$ versus the cylinder circumference $L_y$, with $L_y = 3, 6, 9$. For $J_2 = K_2 = 0$, the data are extrapolated to zero in the quadratic fitting. For nonzero $J_2, K_2$, even a very small number $J_2 = K_2 = 0.00035$, remarkably the linear extrapolation of the RC6 and RC9 data still leads to a nonzero result, indicating a finite three-sublattice order.}
\label{order}
\end{figure}

\section{Model and method}
\label{sec:method}

We perform the DMRG simulations with spin-rotational SU(2) symmetry~\cite{white1992, mcculloch2002} for the system on the square lattice and we consider the SU(3) model by choosing the same bilinear and biquadratic couplings for each bond, as shown in Fig.~\ref{lattice}.
We set the NN couplings $J_1=K_1=1.0$ as the energy scale, and consider additional NNN couplings $J_2=K_2$ along the $e_{x+y}$ direction (see Fig.~\ref{lattice}). 
When $J_2=K_2=1.0$, the system is equivalent to the isotropic triangular-lattice SU(3) model.

In the DMRG simulation, we study the rectangular cylinder (RC) geometry, as shown in Fig.~\ref{lattice}.
This cylindrical geometry has the periodic boundary conditions along the $y$ direction $e_y$ and the open boundaries in the $x$ direction $e_x$. We denote the cylinder as RC$L_y$-$L_x$ ($L_y$ and $L_x$ are the numbers of sites along the two directions, respectively). 
Since the nematic spin liquid and the three-sublattice order states have the dominant structure factors of magnetic correlation functions at ${\bf q} = (\pi, 2\pi/3)$~\cite{hunsl} and $(2\pi/3, 2\pi/3)$~\cite{lauchli2006_2,bauer2012} respectively, we mainly study the systems with $L_y = 6, 9$ and $L_x$ up to $36$, in order to be compatible with both two- and three-sublattice structures. 
We keep up to $6000$ SU(2) multiplets, with the truncation errors smaller than $10^{-5}$. 
For $J_2 = K_2 = 0$, we also study a $L_y = 12$ cylinder by keeping up to $8000$ SU(2) states with the truncation error around $10^{-4}$.

\section{Numerical Results}
\label{sec:results}

In order to detect the three-sublattice magnetic order, one can compute both spin and quadrupolar correlation functions $\langle {\bf S}_{i} \cdot {\bf S}_{j} \rangle $ and $\langle {\bf Q}_{i} \cdot {\bf Q}_{j} \rangle$. 
The quadrupolar tensor operator ${\bf Q}_i$ is defined as 
$( (Q_i^{xx}-Q_i^{yy})/2, (2Q_i^{zz}-Q_i^{xx}-Q_i^{yy})/(2\sqrt{3}), Q_i^{xy}, Q_i^{yz}, Q_i^{xz})$ with $Q_i^{\alpha \beta} = S_i^\alpha S_i^\beta + S_i^\beta S_i^\alpha-\tfrac{4}{3} \delta_{\alpha \beta}$ \mbox{($\alpha,\beta\!=\!x,y,z$)}~\cite{blume1969,toth2010}. 
By definition, the quadrupolar correlation operator ${\bf Q}_{i} \cdot {\bf Q}_{j}$ can be also expressed via bilinear and biquadratic terms as ${\bf Q}_{i}\cdot {\bf Q}_{j}=2({\bf S}_{i}\cdot {\bf S}_{j})^2+{\bf S}_{i}\cdot {\bf S}_{j}-8/3$.

With the computed correlation functions, we can make the Fourier transformations to obtain the spin structure factor as
\begin{equation}
S_{S}({\bf q}) = \frac{1}{N_s}\sum_{i,j} \langle {\bf S}_{i}\cdot {\bf S}_{j} \rangle e^{i{\bf q}\cdot({\bf r}_i-{\bf r}_j)},
\end{equation}
and the quadrupolar structure factor as
\begin{equation}
S_{Q}({\bf q}) = \frac{1}{N_s}\sum_{i,j} \langle {\bf Q}_{i}\cdot {\bf Q}_{j} \rangle e^{i{\bf q}\cdot({\bf r}_i - {\bf r}_j)},
\end{equation}
where the sites $i, j$ are chosen inside the middle region with the size $N_s = L_y \times 2L_y$ in order to avoid open-edge effects and consider both two- and three-sublattice orders~\cite{hunsl}.
Correspondingly, we can further obtain the order parameters $m^2_{S}({\bf q}) = S_{S}({\bf q}) / N_s$ and $m^2_{Q}({\bf q}) = S_{Q}({\bf q}) / N_s$.
Because the magnetic dipolar and quadrupolar correlations consistently describe the three-sublattice order in the SU(3) model simultaneously, we will primarily display the results for the magnetic dipolar correlations below.

\subsection{Confirmation of the spin-liquid state in the nearest-neighbor SU(3) model}

We first reexamine the NN SU(3) model with $J_2=K_2=0$.
In previous numerical studies, the nematic spin liquid and the three-sublattice ordered state appeared to compete with one another.
The two states on finite-size clusters up to $L_y = 9$ seem to have very close energies in DMRG calculations (the DMRG calculations using the boundary-shifted cylinders found the three-sublattice order~\cite{bauer2012}), but the spin liquid state still has the lowest energy on the studied system sizes~\cite{hunsl}.
Here we further study the system with $L_y = 12, L_x = 32$ by pushing the simulation to our limit.

We have mainly checked whether the three-sublattice order would emerge on the largest system size, in case the absent order found in the previous DMRG calculation~\cite{hunsl} was due to finite-size effects.
Therefore, we focused on the spin structure factor $S_{S}({\bf q})$ of the obtained ground state.
If the three-sublattice order eventually wins on this larger size, a sharp peak should be observed at ${\bf q} = (2\pi/3, 2\pi/3)$.
In Fig.~\ref{Ms2}, we show the $S_{S}({\bf q})$ obtained from the spin correlations of the middle $12 \times12$ sites.
It is clear that a round elongated peak appears at ${\bf q} = (\pi, 2\pi/3)$, and symmetry related points, which is highly consistent with the results on the $L_y = 6, 9$ cylinders~\cite{hunsl} and, thus, does not support a three-sublattice order.
We emphasize that although this result for $L_y = 12$ is less converged than for smaller systems as measured by the truncation error, nevertheless the spin structure factor qualitatively agrees well with the previous DMRG results in Ref.~\onlinecite{hunsl} and clearly does {\it not} support the three-sublattice order.

\subsection{Spin correlation function and identification of the three-sublattice order}

We first show our DMRG results by comparing the two extremes, namely by studying the spin dipolar ($m^2_S$) and quadrupolar ($m^2_Q$) order parameters for the spin liquid state at $J_2=K_2=0.0$ and for the three-sublattice order at $J_2=K_2=1.0$ in Fig.~\ref{pd}. In the spin liquid state, both order parameters show a broad horizontally elongated feature at ${\bf q} = (\pi,2\pi/3)$, completely different from the sharp peak at ${\bf q} = (2\pi/3,2\pi/3)$ in the three-sublattice order. Thus, there should be a quantum phase transition between the two phases varying $J_2 = K_2$.

\begin{figure}[!t]
\includegraphics[width = 0.8\linewidth]{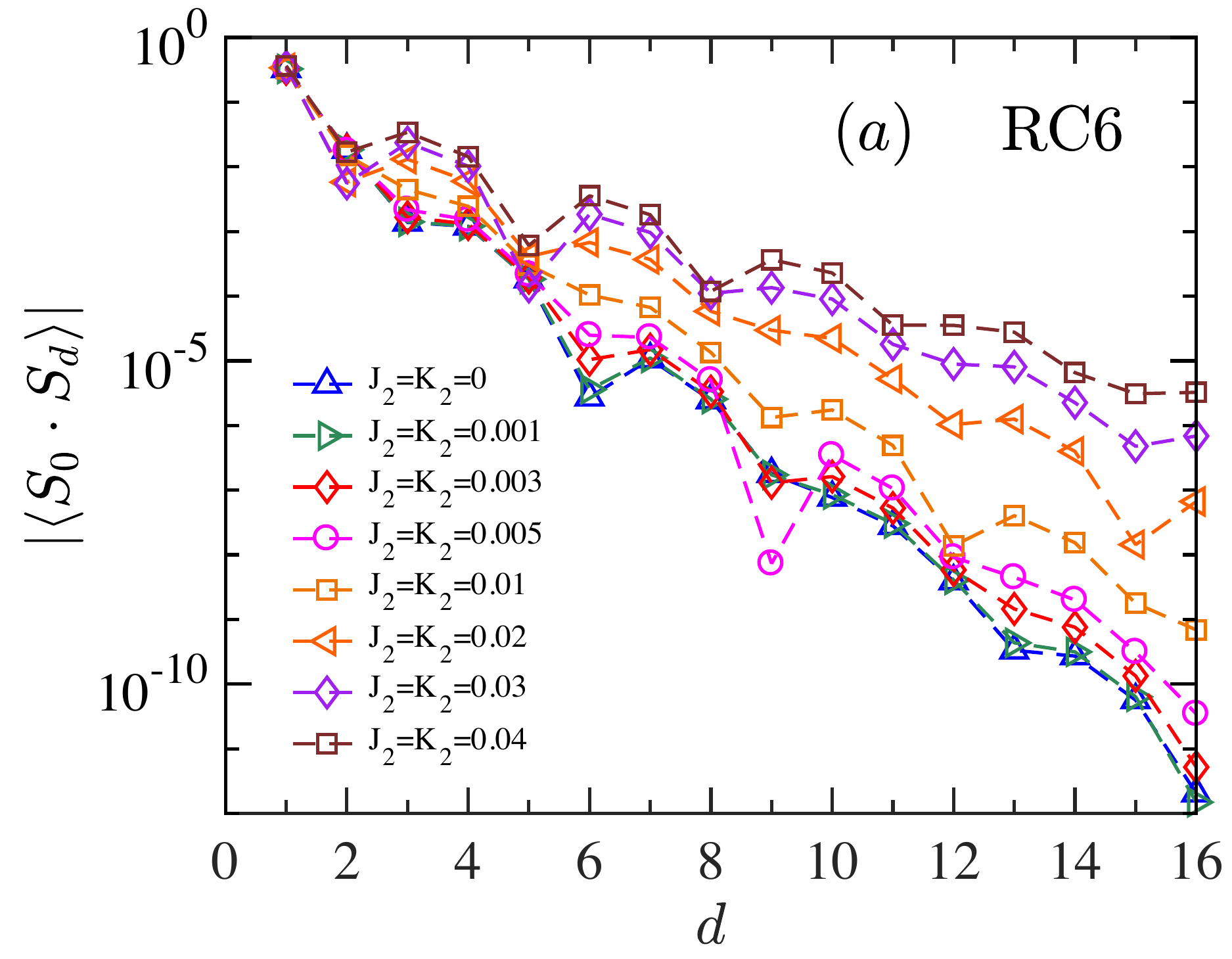}
\includegraphics[width = 0.8\linewidth]{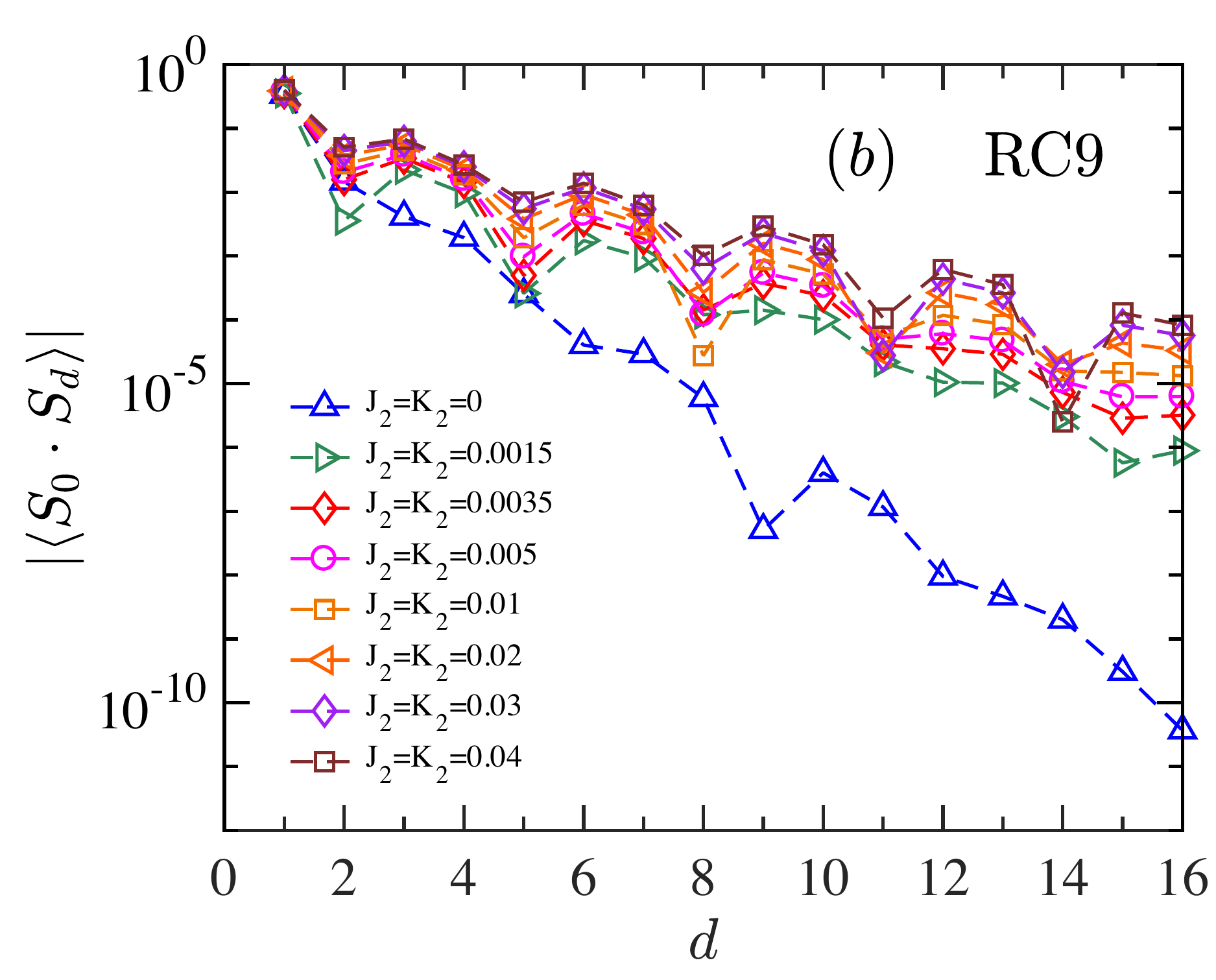}
\caption{$J_2=K_2$ coupling dependence of spin correlation functions in real space.
(a) and (b) are the semilogarithmic plots of spin correlations for different couplings on the RC6 and RC9 cylinders. 
$d$ is the distance between two sites. Particularly for the RC9, the difference between $J_2=K_2=0$ (blue triangles) and rest of the curves is notorious to the eye.}
\label{corr}
\end{figure}

To detect this phase transition, we first study the coupling dependence of spin order parameters.
In Fig.~\ref{order}(a), we show $m^2_S$ at ${\bf q} = (\pi, 2\pi/3)$ and $(2\pi/3, 2\pi/3)$ as a function of $J_2 = K_2$ on the RC6 and RC9 cylinders. 
For $J_2 = K_2 = 0$, the order parameter at $(\pi, 2\pi/3)$ is larger than the one at $(2\pi/3, 2\pi/3)$ { [see the inset of Fig.~\ref{order}(a) for the zoom in of the very small coupling region]}. 
As the interactions $J_2 = K_2$ increase, the order parameter at $(\pi, 2\pi/3)$ decreases slowly. 
On the RC6 cylinder, $m^2_S(2\pi/3, 2\pi/3)$ gradually increases and becomes clearly dominant above $J_2 = K_2 \simeq 0.015$ suggesting already that the critical point is very close to $J_2 = K_2 = 0$.
Moreover, on the RC9 cylinder, $m^2_S(2\pi/3, 2\pi/3)$ shows a sharp enhancement already at even much smaller couplings $J_2 = K_2 \simeq 0.0003$ and then grows gradually, suggesting a possible transition driven by infinitesimal $J_2 = K_2$ couplings. Both lattice sizes with clarity indicate that the transition point is
abnormally close to the $J_2 = K_2 = 0$ point.

To determine whether the three-sublattice order is established or not already in the presence of any nonzero $J_2 = K_2$ couplings, we carry out the finite-size scaling of the order parameter $m^2_S(2\pi/3, 2\pi/3)$ as a function of $1/L_y$ in Fig.~\ref{order}(b).
For $J_2 = K_2 = 0$, the quadratic fitting of $L_y = 3, 6, 9$ data leads to a vanishing order (also see Ref.~\onlinecite{hunsl}).
For nonzero $J_2, K_2$ couplings, even at the smallest couplings we have studied, the linear fitting of $L_y = 6, 9$ data clearly gives a small finite order parameter, which supports that the order appears to be induced by infinitesimal $J_2, K_2$ couplings.

To further support the emergent three-sublattice order at very small $J_2, K_2$ couplings, we investigate the spin correlation function $\langle {\bf S}_i \cdot {\bf S}_j \rangle$ in real space.
We show the semi-logarithmic plots of spin correlations as a function of distance in Fig.~\ref{corr}. 
For the NN SU(3) model, the spin correlations on both RC6 and RC9 decay very fast in an {\it exponential} way $|\langle {\bf S}_i \cdot {\bf S}_j \rangle| \sim e^{-| i - j | / \xi}$ and a small correlation length $\xi \simeq 0.33$, which agrees with the vanishing order identified through finite-size scaling of the order parameter. 
On the RC6 cylinder, spin correlations increase gradually with increasing $J_2, K_2$ (but note the range explored corresponds to very small values of these couplings).
On the wider RC9 cylinder, spin correlations quickly enhance even at the smallest $J_2, K_2$ coupling we have studied, which is consistent with the enhanced order parameter in Fig.~\ref{order}.

\begin{figure}[t]
\includegraphics[width = \linewidth]{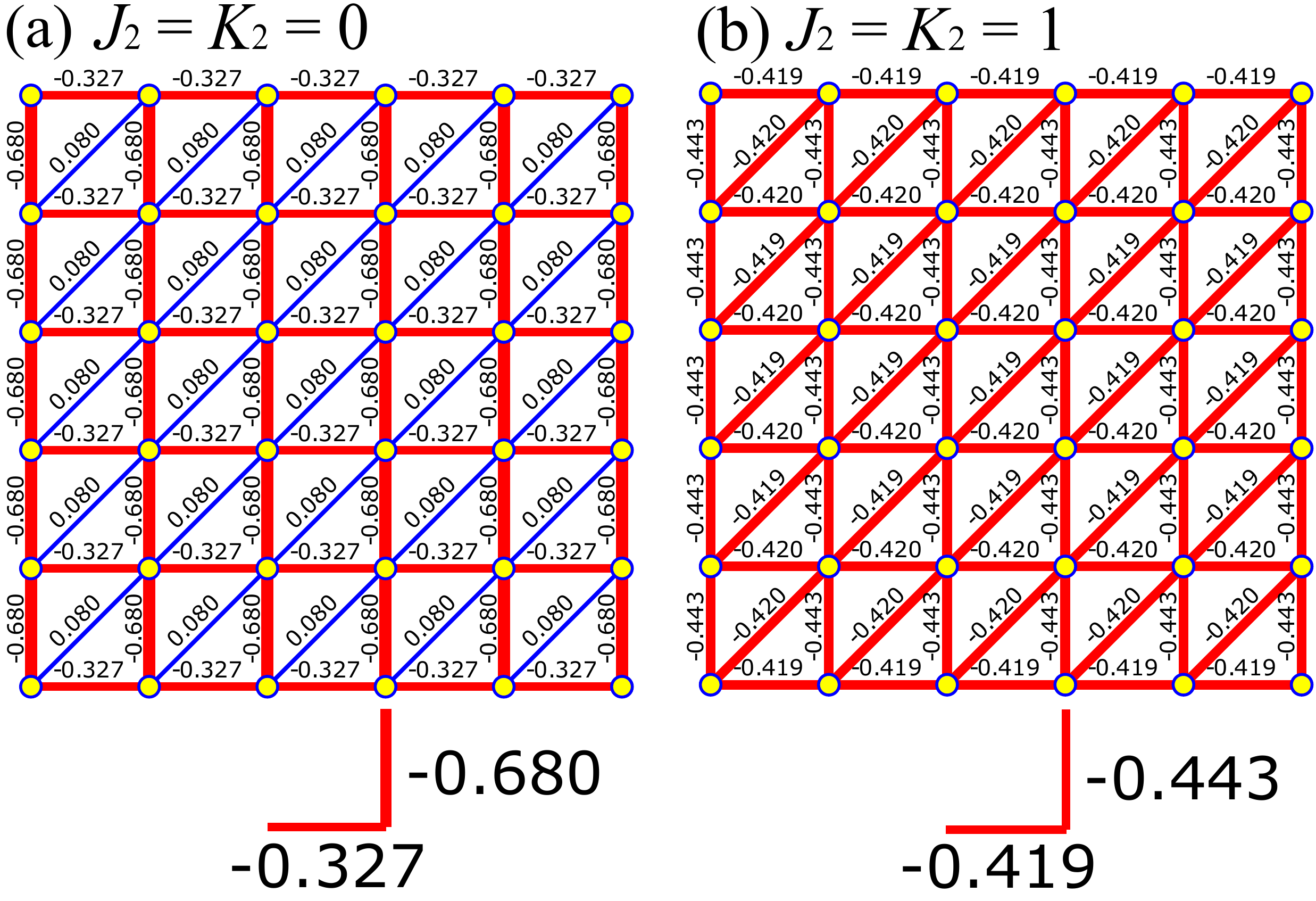}
\caption{Bond correlations $\langle {\bf S}_i \cdot {\bf S}_j \rangle$ of the SU(3) model on the square and triangular lattices. 
(a) are results for the square-lattice model with $J_2 = K_2 = 0$. The blue bonds denote the NNN spin correlation without direct coupling. (b) are results for the triangular-lattice model with $J_2 = K_2 = 1.0$. These bond correlations are measured in the bulk of the RC6 cylinder.  The legends show the bond correlations along the two directions. For the square (a) and triangular (b) models, the absolute values of bond-energy differences are $0.353$ and $0.024$, respectively.}
\label{bond}
\end{figure}

\begin{figure}[t]
\includegraphics[width = 0.8\linewidth]{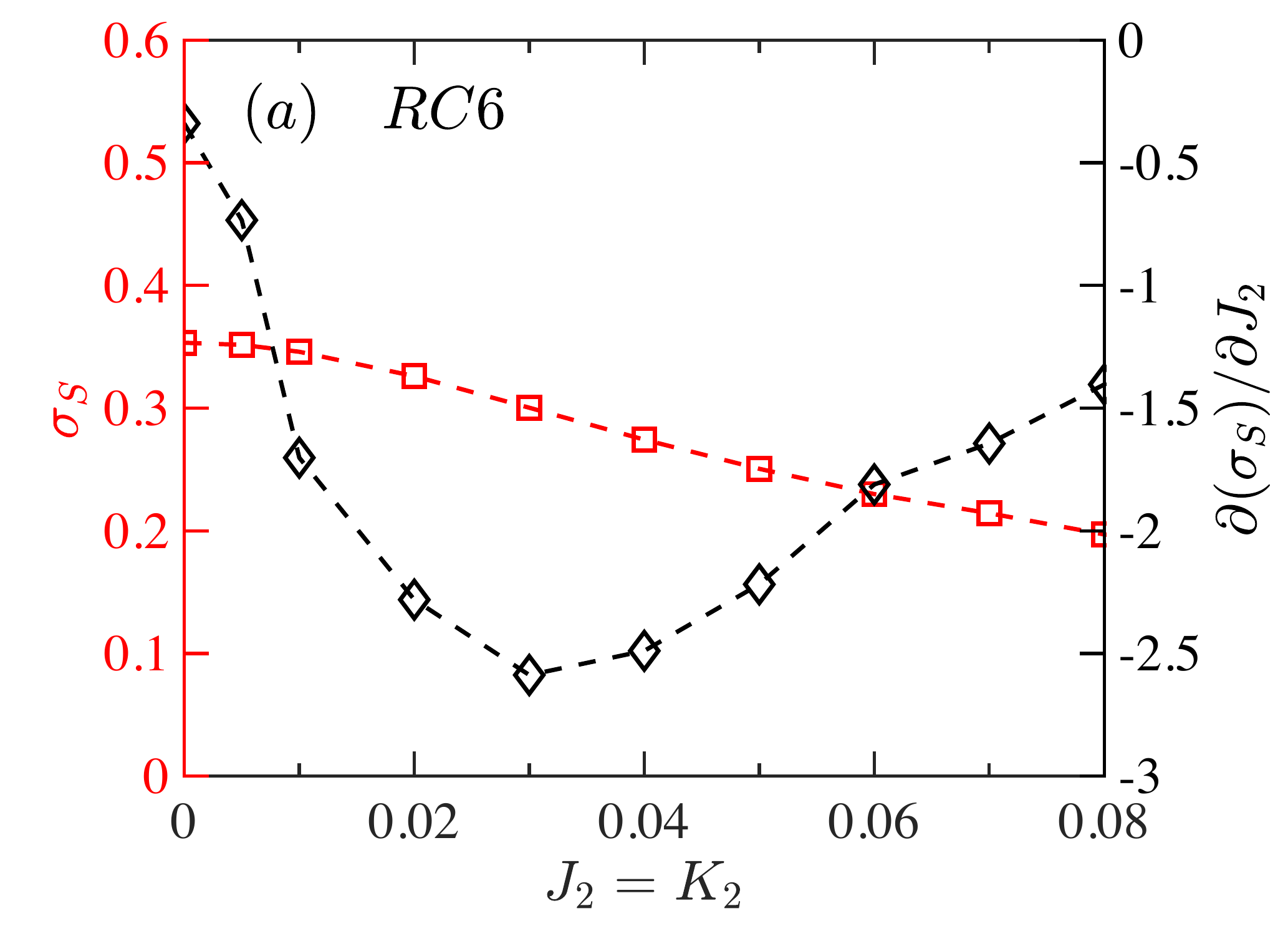}
\includegraphics[width = 0.8\linewidth]{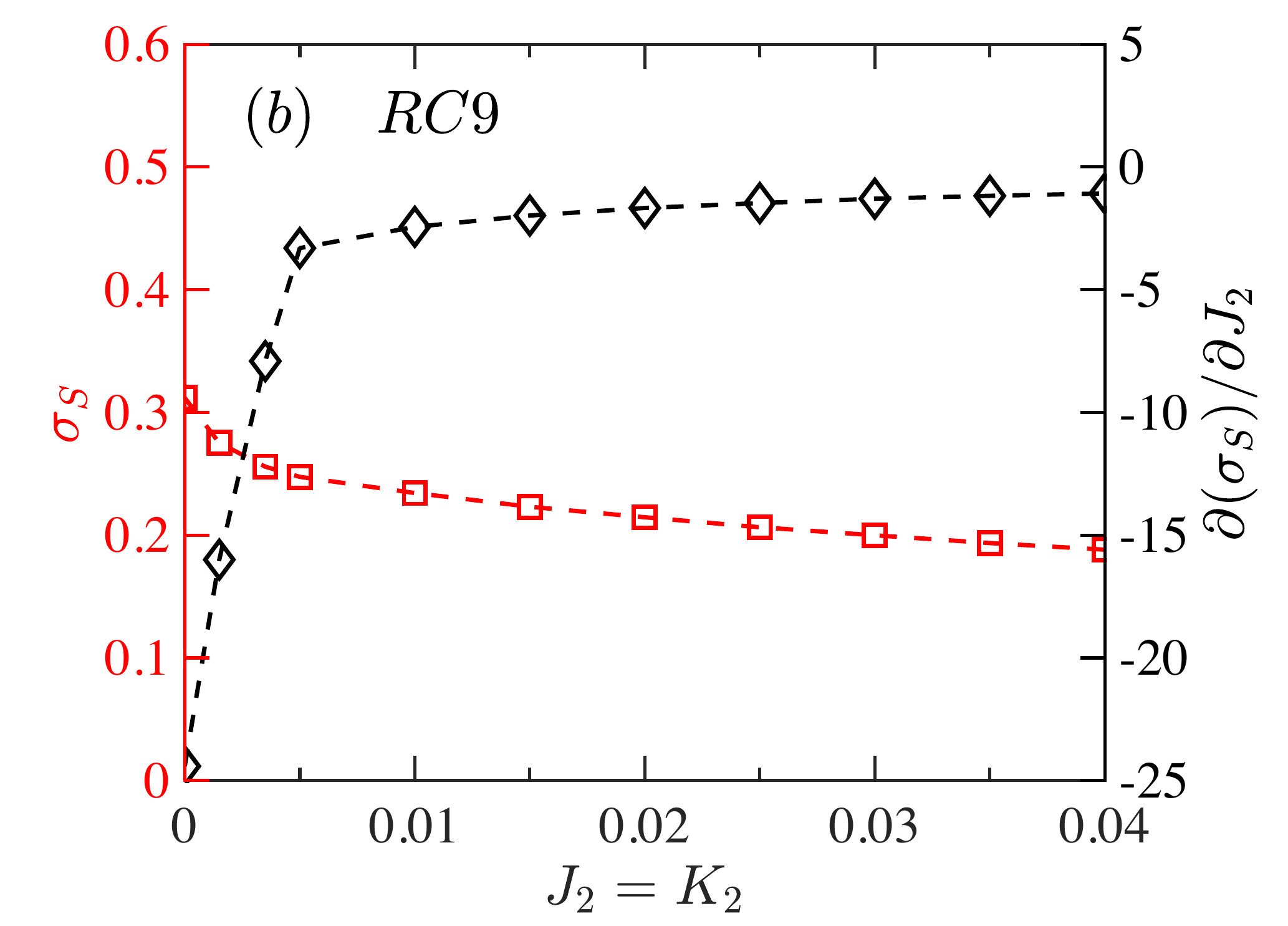}
\caption{Nematic order parameters $\sigma_{S}$ as a function of the couplings $J_2 = K_2$ and its derivative on the RC6 (a) and RC9 (b) cylinders.}
\label{sigma}
\end{figure}

\subsection{Lattice nematic order}
\label{sec:nematic}

Previous DMRG studies of the nematic spin liquid have revealed a spontaneous lattice rotational symmetry breaking with a nonzero lattice nematic order~\cite{b1gb2g,hunematic}, which measures the bond energy difference between the $e_x$ and $e_y$ directions. Such a nematic order parameter can be defined as
\begin{equation}
\sigma_S = \frac{1}{N_m} \sum_i (\langle {\bf S}_i \cdot {\bf S}_{i+\hat{x}} \rangle - \langle {\bf S}_i \cdot {\bf S}_{i+\hat{y}} \rangle),
\end{equation}
where the summations are taken for the $N_m$ sites in the bulk of cylinder.
We show the bond correlations in the dipolar channel in Fig.~\ref{bond}, along the $e_x$, $e_y$, and $e_{x+y}$ directions. 
Notice that although the cylinder geometry used in the DMRG simulation naturally leads to a very small bond nematicity (such geometry-induced nematicity usually decays exponentially with growing system circumference in the quantum states preserving lattice rotational symmetry)~\cite{hunematic, WJHu2020}, it is clear that in the nematic spin liquid state with $J_2 = K_2 = 0$ [see Fig.~\ref{bond}(a)], the bond energies along the $e_x$ ($-0.327$) and $e_y$ ($-0.680$) directions are {\it drastically} different, which have been shown to induce a nonzero $\sigma_S$ after finite-size scaling to the thermodynamic limit~\cite{hunsl}. 
On the other side, in the three-sublattice order phase at $J_2 = K_2 = 1$, the bond energies along the $e_x$ ($-0.419$) and $e_y$ ($-0.443$) directions are quite close to each other [see Fig.~\ref{bond}(b)], consistent with the preserved lattice rotational symmetry of the state.

Because $\sigma_S$ can also be taken as an order parameter to distinguish the two phases in the large-size limit, it is expected that the coupling dependence of $\sigma_S$ on finite-size systems should also provide hints for characterizing the transition.
In Fig.~\ref{sigma}, we show the coupling dependence of $\sigma_S$, as well as its derivative, on the RC6 and RC9 cylinders.
With growing $J_2 = K_2$, $\sigma_S$ continuously decreases.
On the RC6 cylinder, the derivative of $\sigma_S$ has a minimum at $J_2 = K_2 \simeq 0.03$, which is similar to the very small transition coupling found by other procedures before for this cylinder in previous sections.
However, on the larger RC9 cylinder we do not observe any minimum at nonzero $J_2 = K_2$ in the studied parameter region. This result is also consistent with the transition happening at infinitesimal couplings. The fact that the
derivative has a sudden change in slope at $J_2 = K_2 \sim 0.004$ could be used effectively as an upper limit where a transition is possible. But, regardless, it is beyond doubt that the critical point is either at exactly zero,
or at couplings $J_2 = K_2$ abnormally small and likely converging to zero with increasing lattice sizes beyond
the limits of DMRG at present.

In the previous DMRG studies of the SU(3) square-lattice model with additional couplings respecting lattice symmetries~\cite{hunematic}, the identified phase transitions from the nematic spin liquid phase to other symmetry broken phases were found to be highly consistent on the RC6 and RC9 cylinders, suggesting  relatively small finite-size effects.
Here, the sudden transition from the spin liquid to the three-sublattice order when turning on 
$J_2 = K_2$ is clear only after including the wider RC9 system, suggesting stronger finite-size effects in this subtle regime, and displaying the importance of our large-scale calculations.

\section{Summary and Discussion}
\label{sec:sum}

By using large-scale DMRG simulations, we have studied the spin-$1$ SU(3) model on the square-lattice cylinder with not only the NN bilinear-biquadratic couplings $J_1 = K_1 = 1$, but also the additional NNN couplings $J_2 = K_2$ along the $e_{x+y}$ direction (one of the two diagonal directions). 
This system allows a tuning from the square-lattice to the triangular-lattice SU(3) model. 
By calculating spin correlations and different order parameters, we find that the previously found nematic spin liquid state in the NN SU(3) model~\cite{hunsl} may have a transition to the three-sublattice order state at either abnormally small, or even infinitesimal, $J_2, K_2$ couplings.

In previous DMRG studies of this spin-$1$ SU(3) model in Ref.~\onlinecite{hunematic}, both additional $K_2$ and $K_3$ couplings were also considered, but including all the same-distance couplings and thus preserving the lattice symmetries of the square lattice.
In that study, the nematic spin liquid is found to be quite stable for $-0.2 \lesssim K_2 \lesssim 0.3$ or $-0.2 \lesssim K_3 \lesssim 0.1$~\cite{hunematic}.
By sharp contrast, comparing with those previous results the nematic spin liquid in the model studied here seems to be very fragile in the presence of the added diagonal couplings.
We conjecture that the key reason of the fragility of this spin liquid is due to the reduced lattice symmetry in presence of these couplings. 
The symmetry-reduced lattice seems to be less compatible with a two-sublattice structure, which thus may strongly suppress two-sublattice fluctuations and lead to the fast emergence of a three-sublattice order. 
Therefore, in order to stabilize the nematic spin liquid in the spin-1 square-lattice SU(3) model, we believe that the key conceptual aspect is to balance two- and three-sublattice fluctuations.
We also remark that the stabilization of the three-sublattice order at either abnormally small or even infinitesimal diagonal couplings indicates the extremely strong competition of states in the square-lattice SU(3) model. To precisely determine the nature of the ground state, larger system sizes beyond DMRG calculations should be explored in future studies, which may be achieved by recent developments in the finite PEPS simulations~\cite{Liu2021, Liu2020}.

\section*{Acknowledgments}
The authors thank Ling-Fang Lin for her valuable help in the early stages of the project.
X.T.Z. and S.S.G. were supported by the National Natural Science Foundation of China (Grants No. 11874078 and No. 11834014).
E.D. and W.J.H. were supported by the U.S. Department of Energy (DOE), Office of Science, Basic Energy Sciences (BES),  Materials Science and Engineering Division.
Part of the computational calculations have been performed on the Extreme Science and Engineering Discovery Environment (XSEDE) supported by NSF under Grant No.\ DMR160057.

\bibliography{sqtri}

\end{document}